\documentclass{article}%
\usepackage{amsfonts}
\usepackage{amsmath}
\usepackage{amssymb}
\usepackage{charter}
\usepackage{graphicx}
\usepackage{cite}%
\providecommand{\U}[1]{\protect \rule{.1in}{.1in}}

\RequirePackage{CJK}
\AtBeginDocument{\begin{CJK*}{GBK}{song}\CJKtilde}
\AtEndDocument{\end{CJK*}}
\begin{document}

\title{Controllable double optical bistability via photon and phonon interaction in a
hybrid optomechanical system}
\author{Zhen Wang$^{1\ast}$, Cheng Jiang$^{3}$, Yong He$^{4}$, Chang-Ying Wang$^{1}$,
Xiao-Lei Yin$^{1}$,
\and Heng-Mei Li$^{1}$, and Hong-Chun Yuan$^{2}$\\$^{1}${\small School of Science, Changzhou Institute of Technology, Changzhou,
213032, China}\\$^{2}${\small School of Electrical and information Engineering, Changzhou
Institute of Technology, Changzhou, 213032, China}\\$^{3}${\small School of Physics and Electrical Engineering, Huaiyin Normal
University, Huai'an, 223001, China}\\$^{4}${\small School of Mathematics and Physics, Changzhou University,
Changzhou, 213164, China}\\$^{\ast}${\small Corresponding author: wangzhen1712@163.com}}
\maketitle

\begin{abstract}
The optical bistability have been studied theoretically in a multi-mode
optomechanical system with two mechanical oscillators independently coupled to
two cavities in addition to direct tunnel coupling between cavities. It is
proved that the bistable behavior of mean intracavity photon number in the
right cavity can be tuned by adjusting the strength of the pump laser beam
driving the left cavity. And the mean intracavity photon number is relatively
larger in the red sideband regime than that in the blue sideband regime.
Moreover, we have shown that the double optical bistability of intracavity
photon in the right cavity and the two steady-state positions of mechanical
resonators can be observed when the control field power is increased to a
critical value. Besides, the critical values for observing bistability and
double bistability can be tuned by adjusting the coupling coefficient between
two cavities and the coupling rates between cavities mode and mechanical mode.

\end{abstract}


(Some figures may appear in colour only in the online journal)

\section{Introduction}

Optomechanical systems, in which light field interacts with mechanical
resonators via radiation pressure, is a fast developing area in quantum optics
domain due to the promising applications in ultrasensitive
detecting\cite{1,2,3}, mirror ground-state cooling\cite{4,5,6,7}, and quantum
entanglement generation\cite{8,9}. Particularly, the typical optomechanical
cavities with one or more movable mirrors are attracting increasingly
attention\cite{10,11,12,13} as the hybrid optomechanical coupling is well
exploited in the realization of quantum information processing\cite{14,15,16}
and in the observation of macroscopic quantum behavior\cite{17}.

Among the plenty of nonlinear phenomena in cavity optomechanical systems, the
optical bistability which is characterized by the intracavity mean photon
number have been investigated in different optomechanical
systems\cite{18,19,20} and under the effect of different type of
interactions\cite{21,22,23}. Optical bistability denotes that it is possible
to deliver two different outputs for an applied power to the system. The
essence of observing bistability in optomechanical systems is referred to the
nonlinear nature of coupling between radiation pressure and mechanical
oscillations\cite{24,25}. It is mentionable that the optical bistability can
be controlled by a strong laser field and was first experimentally observed by
Dorsel et al.\cite{26}. And subsequently, the bistable behavior of the mean
intracavity photon number in optomechanical systems with Bose-Einstein
condensate\cite{27,28}, ultracold atoms\cite{29,30}, and quantum well\cite{31}
has also been extensively studied. Particularly, optical bistability and
dynamic instability are experimentally observed and studied in a system
consisting of three-level L-type rubidium atoms in an optical ring cavity. The
bistable behavior and self-pulsing frequency are experimentally manipulated by
changing the controlling and cavity field parameters\cite{32}. The nonlinear
phenomenon of optical bistability inside a ring resonator formed with a
silicon-waveguide nanowire has been theoretically analyzed and an exact
parametric relation connecting the output intensity to the input intensity has
been derived\cite{33}. Optical bistability in coupled optomechanical cavities
in the presence of Kerr effect has been studied, and it is found that the
atomic medium has a deep effect on bistable behavior of intracavity intensity
for the optomechanical cavity, and meanwhile a critical value for the Kerr
coefficient to observe bistability in intracavity intensity for the
optomechanical cavity is determined\cite{34}.

Moreover, certain works concerning the multi-mode optomechanical system have
been extensively investigated. Specifically, in the three-mode optomechanical
systems where one mechanical mode is optomechanically coupled to two linearly
coupled optical modes simultaneously, the possibility of optical nonreciprocal
response have been demonstrated, and the system can be used as a three-port
circulator for two optical modes and one mechanical mode\cite{35}.
Additionally, it has demonstrated that the nonreciprocal conversion between
microwave and optical photons in an electro-optomechanical system where a
microwave mode and an optical mode are coupled indirectly via two
nondegenerate mechanical modes, and the electro-optomechanical system can also
be used to construct a three-port circulator for three optical modes with
distinctively different frequencies by adding an auxiliary optical mode
coupled to one of the mechanical modes\cite{36}. In\cite{37}, Malz et al. have
realized the implementation of phase-preserving and phase-sensitive
directional amplifiers for microwave signals in an electromechanical setup
comprising two microwave cavities and two mechanical resonators. And it is an
important step towards flexible and on-chip integrated nonreciprocal
amplifiers of microwave signals. Regarding to the preceding efforts, in this
paper we intend to study the bistable behavior of the hybrid optomechanical
systems consisting of two optomechanical cavities coupled via two mechanical
resonators in addition to direct tunnel coupling. Next, we shall theoretically
investigate the bistability of the intracavity photon number and mechanical
steady-state positions in the hybrid optomechanical systems. The bistable
behavior of the steady-state photon number and the mechanical steady-state
positions can be effectively adjusted by the power of pump field, the coupling
between cavity and mechanical resonator, and the detuning between cavity and
pump field.

The remaining content of the work is arranged as follows: in Sec.2,\ we will
introduce the theoretical mode, the Hamiltonian, and the nonlinear quantum
Langevin equations associated to the Hamiltonian. In Sec.3, the numerical
results are presented and discussed in detail according to the steady state
solution of the Langevin equations. The last section is devoted to make conclusions.

\section{Description of the system}

\subsection{Theoretical Mode and Hamiltonian}

The system under consideration is shown in Fig.1, which is consisted by two
optical cavities coupled via two mechanical resonators respectively in
addition to direct tunnel coupling\cite{36,37}.

\begin{figure}[ptb]
\label{Figure1}
\centering \includegraphics[width=9cm]{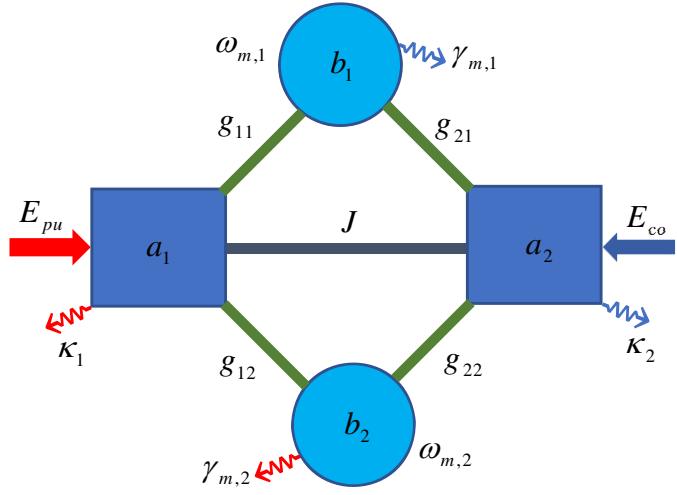}\caption{The schematic
diagram of multi-mode optomechanical systems where the two optical cavity
modes, $a_{1}$ and $a_{2}$ (with decay rate $\kappa_{1}$ and $\kappa_{2}$,
respectively), are linearly coupled to each other via a common waveguide with
coupling coefficient $J$, and two mechanical modes $b_{1}$ with frequency
$\omega_{m,1}$ and $b_{2}$ with frequency $\omega_{m,2\text{ }}$(with damping
rate $\gamma_{m,1}$ and $\gamma_{m,2}$, respectively), are optomechanically
coupled to the two optical modes, respectively. The left cavity is driven by a
pump beam $E_{pu}$ and the right cavity is driven by a control beam $E_{co}$.}%
\end{figure}

In order to investigate the optical response of the hybrid system, we assume
that the two cavity modes are driven simultaneously by a strong pump field
with frequency $\omega_{pu}$ and a weaker control field with frequency
$\omega_{co}$, respectively. The Hamiltonian describing the system is given as%

\begin{align}
H  &  =\sum_{l=1}^{2}\hslash \left(  \omega_{l}a_{l}^{\dagger}a_{l}%
+\omega_{m,l}b_{l}^{\dagger}b_{l}\right)  -\sum_{k,l=1}^{2}\hslash g_{kl}%
a_{k}^{\dagger}a_{k}\left(  b_{l}^{\dagger}+b_{l}\right) \nonumber \\
&  -\hslash J\left(  a_{1}^{\dagger}a_{2}+a_{1}a_{2}^{\dagger}\right)
+H_{dr}, \label{1}%
\end{align}
where $a_{l}^{\dagger}\left(  a_{l}\right)  $ and $b_{l}^{\dagger}\left(
b_{l}\right)  $ are the creation (annihilation) operator of the $l$th cavity
mode and mechanical resonator, respectively. The $\omega_{l}$ and
$\omega_{m,l}$ are the frequencies of the $l$th cavity mode and mechanical
resonator mode. The $g_{kl}$\ represents the coupling rate between the $k$th
optical cavity and the $l$th mechanical oscillator. And $J$ is the photon
tunneling amplitude through the two central mirror. The last term $H_{dr}$
describes the interaction between the driving fields and the optomechanical
system: A strong pump field $E_{pu}$ of frequency $\omega_{pu}$ and a weak
probe field of frequency $\omega_{pr}$ are simultaneously employed to drive
the cavity mode $a_{1}$, and another strong control field $E_{co}$ of
frequency $\omega_{co}$ is applied to drive the cavity $a_{2}$, i.e.,%

\begin{equation}
H_{dr}=i\hslash \sqrt{\kappa_{e1}}E_{pu}e^{-i\omega_{pu}t}a_{1}^{\dagger
}+i\hslash \sqrt{\kappa_{e2}}E_{co}e^{-i\omega_{co}t}a_{2}^{\dagger
}+\text{H.c.}, \label{2}%
\end{equation}
where $E_{pu}$ and $E_{co}$ are the laser amplitude, i.e., $\left \vert
E_{pu}\right \vert =\sqrt{\frac{2\kappa_{1}P_{pu}}{\hslash \omega_{pu}}}$ and
$\left \vert E_{co}\right \vert =\sqrt{\frac{2\kappa_{2}P_{co}}{\hslash
\omega_{co}}}$, with $P_{pu}$ the power of laser corresponding to pump field,
$P_{co}$ the power of laser corresponding to control field, and $\kappa_{l}$
the cavity decay rate associated with the $l$th cavity mode $a_{l}$ ($l=1,2$).

In the rotating frame at the pump frequency $\omega_{pu}$, the Hamiltonian of
the hybrid optomechanical system reads:%

\begin{align}
H^{\prime}  &  =\sum_{l=1}^{2}\hslash \left(  \Delta_{l}a_{l}^{\dagger}%
a_{l}+\omega_{m,l}b_{l}^{\dagger}b_{l}\right)  -\sum_{l,k=1}^{2}\hslash
g_{lk}a_{l}^{\dagger}a_{l}\left(  b_{k}^{\dagger}+b_{k}\right) \nonumber \\
&  -\hslash J\left(  a_{1}^{\dagger}a_{2}+a_{1}a_{2}^{\dagger}\right)
+H_{dr}, \label{3}%
\end{align}
where $\Delta_{1}=\omega_{1}-\omega_{pu}$, $\Delta_{2}=\omega_{2}-\omega_{co}$.

\subsection{Heisenberg-Langevin Equations}

According to the Heisenberg-Langevin equations of motion\cite{38}, after
introducing the corresponding damping and noise terms to the equations of
motion associated with the Hamiltonian in Eq.(\ref{3}), one can gets the
following set of nonlinear equations which read as%
\begin{align}
\frac{da_{1}}{dt}  &  =-\left(  i\Delta_{1}+\frac{\kappa_{1}}{2}\right)
a_{1}+ig_{11}a_{1}X_{1}+ig_{12}a_{1}X_{2}+iJa_{2}\nonumber \\
&  +\sqrt{\kappa_{e1}}E_{pu}+\sqrt{2\kappa_{1}}a_{in,1},\nonumber \\
\frac{da_{2}}{dt}  &  =-\left(  i\Delta_{2}+\frac{\kappa_{2}}{2}\right)
a_{2}+ig_{21}a_{2}X_{1}+ig_{22}a_{2}X_{2}+iJa_{1}\nonumber \\
&  +\sqrt{\kappa_{e2}}E_{co}+\sqrt{2\kappa_{2}}a_{in,2},\nonumber \\
\frac{d^{2}X_{1}}{dt^{2}}  &  =-\gamma_{m,1}\frac{dX_{1}}{dt}-\left(
\omega_{m,1}\right)  ^{2}X_{1}+2\omega_{m,1}g_{11}a_{1}^{\dagger}%
a_{1}\nonumber \\
&  +2\omega_{m,1}g_{21}a_{2}^{\dagger}a_{2}+\xi_{1},\nonumber \\
\frac{d^{2}X_{2}}{dt^{2}}  &  =-\gamma_{m,2}\frac{dX_{2}}{dt}-\left(
\omega_{m,2}\right)  ^{2}X_{2}+2\omega_{m,2}g_{12}a_{1}^{\dagger}%
a_{1}\nonumber \\
&  +2\omega_{m,2}g_{22}a_{2}^{\dagger}a_{2}+\xi_{2}, \label{4}%
\end{align}
where $X_{l}=a_{l}^{\dagger}+a_{l}$ and $a_{in,1\text{ }}$and $a_{in,2\text{
}}$are the quantum vacuum fluctuations of the cavity mode $a_{1\text{ }}$and
$a_{2\text{ }}$, which are fully characterized by the correlation
$\left \langle a_{in,l}\left(  t\right)  a_{in,l}^{\dagger}\left(  t^{\prime
}\right)  \right \rangle =\delta \left(  t-t^{\prime}\right)  $, $\left \langle
a_{in,l}^{\dagger}\left(  t\right)  \right.  $ $\  \left.  a_{in,l}\left(
t^{\prime}\right)  \right \rangle =0$, $\xi_{l}$ is the Brownian stochastic
force with zero mean value that obeys the correlation function $\left \langle
\xi_{l}\left(  t\right)  \xi_{l}\left(  t^{\prime}\right)  \right \rangle
=\frac{\gamma_{m,l}}{\omega_{m,l}}\int \frac{d\omega}{2\pi}\omega
e^{-i\omega \left(  t-t^{\prime}\right)  }\left[  \coth \left(  \frac
{\hslash \omega}{2k_{B}T_{l}}\right)  +1\right]  $ with $l=1,2$. Here $k_{B}$
is the Boltzmann constant and $T_{l}$ is the temperature of the reservoir of
the $l$th mechanical resonator. The cavity modes decay at the rate $\kappa
_{l}$ and are affected by the input vacuum noise operator $a_{in,l}$ with zero
mean value, the mechanical mode is affected by a viscous force with damping
rate $\gamma_{m,l}$ and by a Brownian stochastic force with zero mean value
$\xi$\cite{39}. To study the bistability of the presented system, we are
interested in the steady-state solutions to the Eqs.(\ref{4}). And the mean
intracavity photon number $n_{p,l}=$ $\left \vert a_{l,s}\right \vert ^{2}$
($l=1,2$) can be determined by the coupled equations as following%
\begin{align}
n_{p,1}  &  =\frac{\kappa_{e1}\left(  E_{pu}\right)  ^{2}+J^{2}n_{p,2}%
}{\left(  \frac{\kappa_{1}}{2}\right)  ^{2}+\left(  \Delta_{1}-g_{11}%
X_{1,s}-g_{12}X_{2,s}\right)  ^{2}},\nonumber \\
n_{p,2}  &  =\frac{\kappa_{e2}\left(  E_{co}\right)  ^{2}+J^{2}n_{p,1}%
}{\left(  \frac{\kappa_{2}}{2}\right)  ^{2}+\left(  \Delta_{2}-g_{21}%
X_{1,s}-g_{22}X_{2,s}\right)  ^{2}}, \label{5}%
\end{align}
where $X_{1,s}$ and $X_{2,s}$ are the mechanical steady-state positions which
are related to mean intracavity photon number:%
\begin{align}
X_{1,s}  &  =\frac{2}{\omega_{m,1}}\left(  g_{11}n_{p,1}+g_{21}n_{p,2}\right)
,\nonumber \\
X_{2,s}  &  =\frac{2}{\omega_{m,2}}\left(  g_{12}n_{p,1}+g_{22}n_{p,2}\right)
. \label{6}%
\end{align}

\section{Results and discussion}

In this section, we shall numerically investigate the bistable behavior\ when
the two cavities are respectively coupled by two mechanical resonators in
addition to direct tunnel coupling. We consider an experimentally realized
optomechanical system. Thus we choose the parameters similar to those
in\cite{40} $\omega_{pu}=2\pi \times205.3$THz, $\omega_{co}=2\pi \times
194.1$THz, $\omega_{m,1}$($\omega_{m,2}$)$=2\pi \times2.0$GHz, $\kappa_{1}%
=2\pi \times520$MHz, $\kappa_{2}=2\pi \times1.73$GHz, $\kappa_{e1}=2\pi
\times0.26$MHz, and $\kappa_{e2}=2\pi \times8.0$MHz.

\subsection{Intracavity photon number}

Firstly, we investigate the optical bistability in the left cavity by
adjusting the strengths of pump beam. The variations of the intracavity photon
in the left cavity and right cavity, respectively, versus the left cavity-pump
field detuning $\Delta_{1}$ for different pump strengths are shown in Fig.2(a)
and Fig.2(b). As can be seen in Fig.2(a), the curve is a nearly Lorentzian
peak when the strength of the left pump beam is lower; however, when the
strength increases above a critical value, the system exhibits obvious
bistable behavior, as shown in the curves for a range of values of the driving
laser strength. Specifically, the initially nearly Lorentzian resonance curve
becomes clear asymmetric in\ the pump beam strength $P_{pu}=0.10\mu$W and
$P_{pu}=0.20\mu$W. In this case, the coupled cubic equation(\ref{5}) for the
mean intracavity photon number yield three real roots. The largest and
smallest roots are stable, and the middle one is unstable, which is
represented by the dashed lines in Fig.2(a). Furthermore, we can see that the
larger cavity pump detuning is necessary to observe the optical bistable
behavior with the increasing pump beam strength. The mean intracavity photon
number $n_{p,2}$ in the right cavity as a function of the left cavity-pump
beam detuning $\Delta_{1}$ is plotted in Fig.2(b). In this case, different
from the behavior in the left cavity, the mean intracavity photon number all
smaller than a certain value. When the strength of the pump beam is lower, the
curve is a nearly Lorentzian dip. However, when the strength increases above a
critical value, the system exhibits obvious bistable behavior for a range of
values of the driving laser strength, as shown in the curves for\ $P_{pu}%
=0.10\mu$W and $0.20\mu$W, where the initially nearly Lorentzian curve becomes
apparently asymmetric.

\begin{figure}[ptb]
\label{Figure2}
\centering \includegraphics[width=9cm]{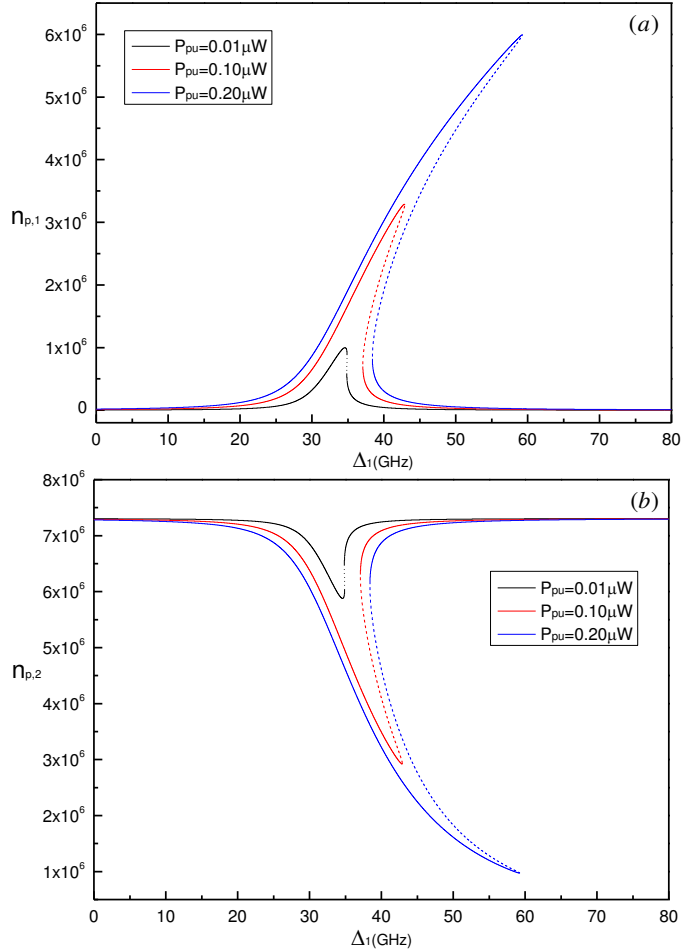}\caption{The mean
intracavity photon number in (a) the left cavity $n_{p,1}$ and (b) the right
cavity $n_{p,2}$ as a function of $\Delta_{1}$ with $J=$ $2\pi \times0.09$GHz,
$g_{11}=2\pi \times850$kHz, $g_{12}=2\pi \times860$kHz, $g_{21}=2\pi \times
400$kHz, $g_{22}=2\pi \times405$kHz, $\Delta_{2}=2\pi \times2.0$GHz, and
$P_{co}=0.03\mu$W for different values of $P_{pu}$: $P_{pu}=0.01\mu$W,
$0.10\mu$W, and $0.20\mu$W, respectively.}%
\end{figure}

The optical bistability in the two cavities can be equivalently seen from the
hysteresis loop for the mean intracavity photon number versus the pump power
for a range of control power, as shown in Fig.3 and Fig.4. The solid and
dashed curves are, respectively, corresponding to stable and unstable
solutions. It is shown in Fig.3 that the mean intracavity photon number
$n_{p,1}$ initially lies in the lower stable branch (corresponding to the
smallest root). When the input pump power is gradually increased, the
intracavity intensity in the left cavity initially scans the lower branch of
the curve. When it arrives at the end of the lower branch, i.e., the first
critical value, then it jumps to the upper stable branch. Therefore,
increasing of input pump power shall leads to the increasing of intracavity
photon number. After jumping to the upper branch, the intracavity intensity
starts decreasing if the input pump power is decreased, but it still follows
the upper stable branch. When the intracavity intensity reaches to the second
critical point, it will jump down to the lower stable branch. Moreover, with
the strengthening of control power, the pump power needed to observe the
optical bistability is relatively lower.

The bistable behavior in the right cavity can be alternatively seen from the
hysteresis loop for the mean intracavity photon number $n_{p,2}$ versus the
pump power curve shown in Fig.4. Different from the case in the left cavity,
as the power of control field is increased, it is interesting to notice that
there emerges the double bistability in the right cavity. Specifically, the
intracavity intensity in the right cavity $n_{p,2}$ initially scans the first
lower branch of the curve when the input pump power is gradually increased.
When it arrives at the end of the first lower branch, then it jumps to the
first upper stable branch. After jumping to the upper branch, the intracavity
intensity starts to increase if the input pump power is further increased;
however, if the input pump power is further increased, then the intracavity
intensity in the right cavity will jumps to the second upper stable branch. It
is obvious that, with increasing control power, the second cavity switches
from a bistable regime to a double bistable one. Physical concept behind this
result can be expressed as the follows. A higher value of control power leads
to enhanced nonlinearity in right cavity, and then the increased nonlinearity
of the system shall leads to double bistable behavior in the cavity.

\begin{figure}[ptb]
\label{Figure3}
\centering \includegraphics[width=9cm]{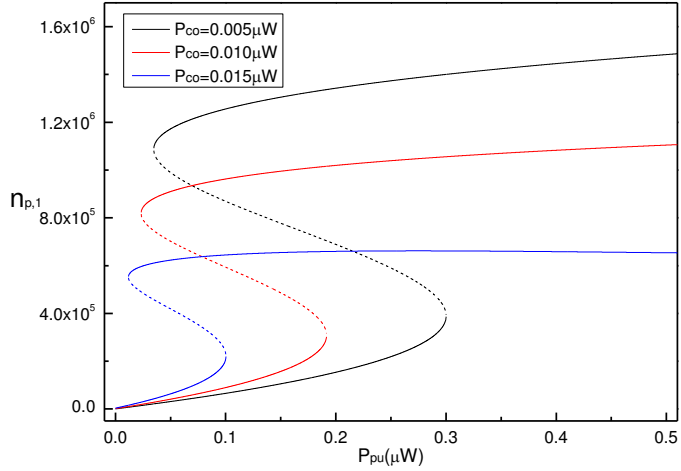}\caption{The mean
intracavity photon number in the left cavity $n_{p,1}$ as a function of
$P_{pu}$ with $J=$ $2\pi \times0.09$GHz, $g_{11}=2\pi \times850$kHz,
$g_{12}=2\pi \times860$kHz, $g_{21}=2\pi \times400$kHz, $g_{22}=2\pi \times
405$kHz, and $\Delta_{1}=\Delta_{2}=2\pi \times2.0$GHz for different values of
$P_{co}$: $P_{co}=0.005\mu$W, $0.010\mu$W, and $0.015\mu$W, respectively.}%
\end{figure}

\begin{figure}[ptb]
\label{Figure4}
\centering \includegraphics[width=9cm]{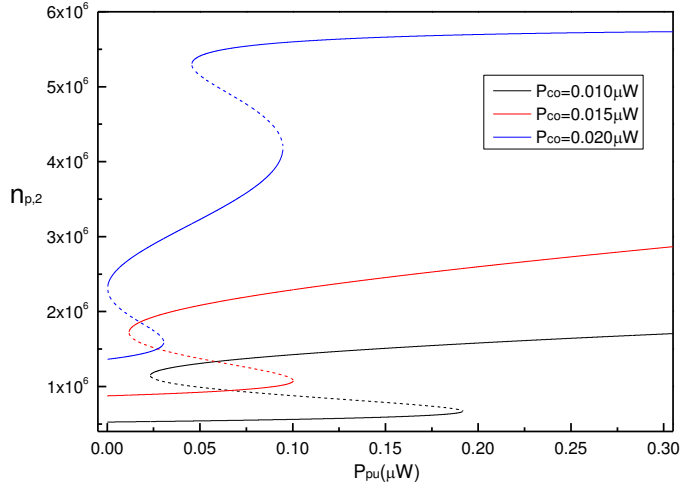}\caption{The mean
intracavity photon number in the right cavity $n_{p,2}$ as a function of
$P_{pu}$ with $J=$ $2\pi \times0.09$GHz, $g_{11}=2\pi \times850$kHz,
$g_{12}=2\pi \times860$kHz, $g_{21}=2\pi \times400$kHz, $g_{22}=2\pi \times
405$kHz, and $\Delta_{1}=\Delta_{2}=2\pi \times2.0$GHz for different values of
$P_{co}$: $P_{co}=0.010\mu$W, $0.015\mu$W, and $0.020\mu$W, respectively.}%
\end{figure}

Bistability in the system also sensitively affected by the coupling rate $J$
between the two cavities, as depicted in Fig.5. In the weak coupling regime,
the pump beam driving the left cavity cannot have enough impact on the photon
numbers in the right cavity via the waveguide, thus the threshold value to
observe bistability is relatively larger compared with the situation in
stronger coupling regime. Thus, with the strengthening of coupling between the
two cavities, the pump intensity needed to observe the optical bistability is
relatively lower. It should be noted that the upper stable branches in the
hysteresis loop of left cavity for different coupling intensities $J$ approach
each other eventually in the larger pump beam intensity. The physical
mechanism lies in the fact that the intracavity photon numbers are not only
affected by the pump field but also affected by the control field through the
tunnel coupling between the two cavities. Thus, the threshold of pump
intensity to observe the optical bistability is relatively lower when there is
stronger coupling. Additionally, the control field has little effect on photon
number in the left cavity in times of stronger pump driving. Thus, it results
in the phenomenon in which the upper stable branches in the hysteresis loop
approach each other in larger pump beam intensity for different coupling
intensities. Meanwhile, the pump intensity required to observe the optical
bistability of the photon number in the right cavity will become lower with
the increase of coupling intensity between the two cavities. This is similar
to the case in the left cavity. The reason is that with the increasing of
tunneling coupling between the two cavities, there are more photons from the
left cavity tunneling to the right cavity. Then it is more likely to observe
the bistable phenomenon in the right cavity, and thus the pump intensity
required to observe the bistability is relatively lower. The photon number of
the right cavity is affected by both the pump field and the control field.
With the increase of coupling strength, the effect of pump field becomes
stronger, which leads to a larger intracavity photon number in the upper
stable branch in the hysteresis. As depicted in Fig.6, in the case of stronger
control field comparing with that in Fig.5, with the increase of strengthening
in coupling rate between the two cavities, it is more easier to observe the
double bistability, for the threshold value of pump power to generate
bistability is lower. The reason of observing optical bistability at lower
pump power in the larger coupling intensity is a result of the enhanced
interaction of control field through the tunnel effect.

\begin{figure}[ptb]
\label{Figure5}
\centering \includegraphics[width=9cm]{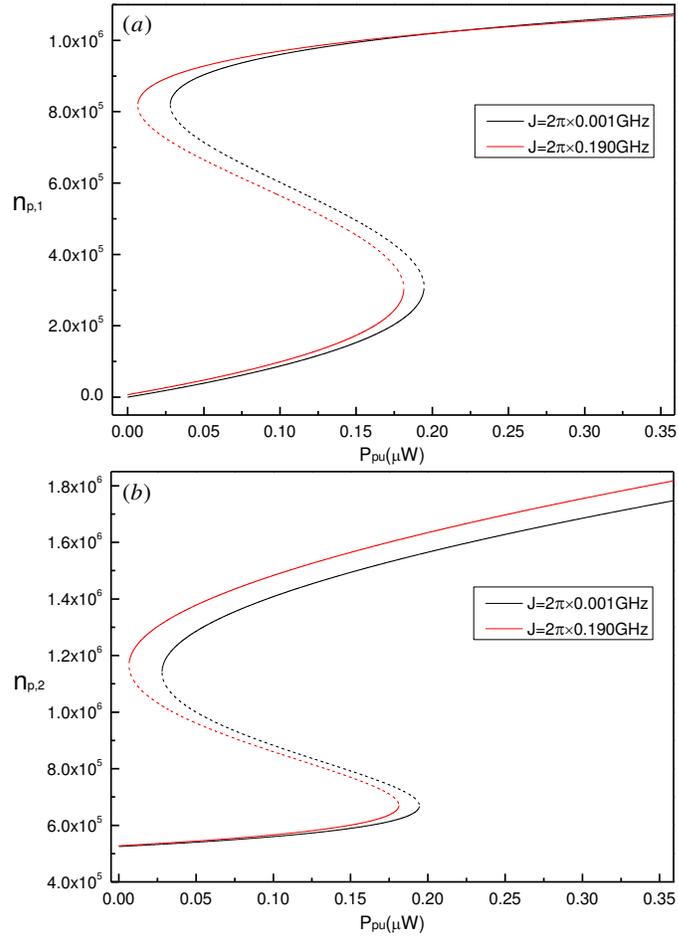}\caption{The mean
intracavity photon number in (a) the left cavity $n_{p,1}$ as a function of
$P_{pu}$ and (b) the right cavity $n_{p,2}$ as a function of $P_{pu}$ with
$P_{co}=0.01$ $\mu$W, $g_{11}=2\pi \times850$kHz, $g_{12}=2\pi \times860$kHz,
$g_{21}=2\pi \times400 $kHz, $g_{22}=2\pi \times405$kHz, $\Delta_{1}=\Delta
_{2}=2\pi \times2.0$GHz for different values of $J$: $J=2\pi \times0.001$GHz and
$2\pi \times0.190$GHz, respectively. }%
\end{figure}

\begin{figure}[ptb]
\label{Figure6}
\centering \includegraphics[width=9cm]{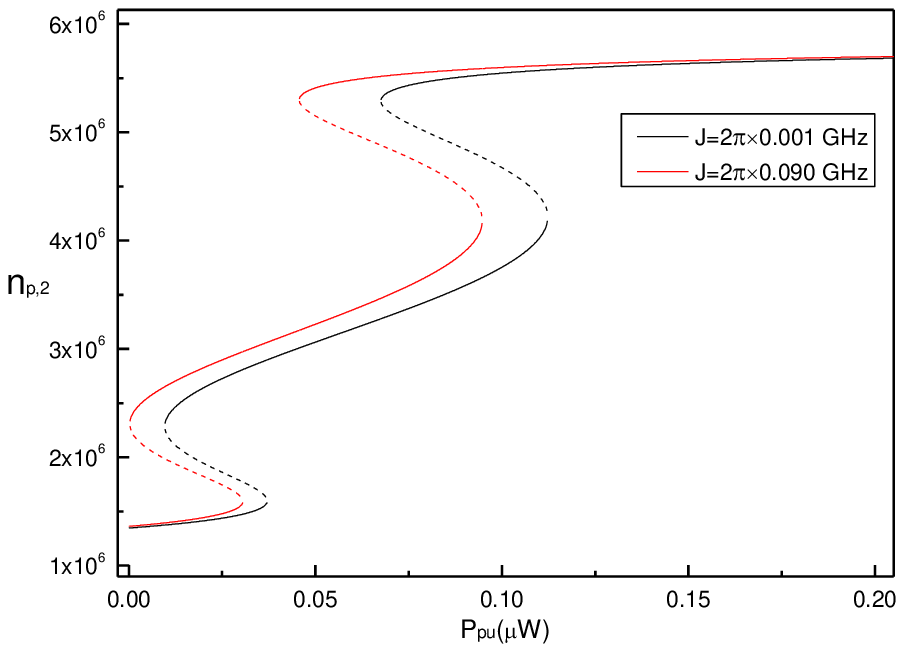}\caption{The mean
intracavity photon number in the right cavity $n_{p,2}$ as a function of
$P_{pu}$ with $P_{co}=0.02$ $\mu$W, $g_{11}=2\pi \times850$kHz, $g_{12}%
=2\pi \times860$kHz, $g_{21}=2\pi \times400$kHz, $g_{22}=2\pi \times405$kHz,
$\Delta_{1}=\Delta_{2}=2\pi \times2.0$GHz for different values of $J$:
$J=2\pi \times0.001$GHz and $2\pi \times0.090$GHz, respectively. }%
\end{figure}

If the right cavity is pumped on the red sideband, i.e., $\Delta_{2}$ $=$
$\omega_{m,2}$, then it can be seen from Fig.7 that the threshold value to
observe the optical bistability in the left cavity is lower compared with that
in blue sideband, i.e., $\Delta_{2}$ $=$ $-\omega_{m,2}$. However, the upper
stable branches of the hysteresis loop in blue sideband is larger than that in
red sideband. The mean intracavity photon number in right cavity $n_{p,2}$
under the condition of red sideband is manifestly smaller than the value
obtained in blue sideband, as shown in Fig.8.

\begin{figure}[ptb]
\label{Figure7}
\centering \includegraphics[width=9cm]{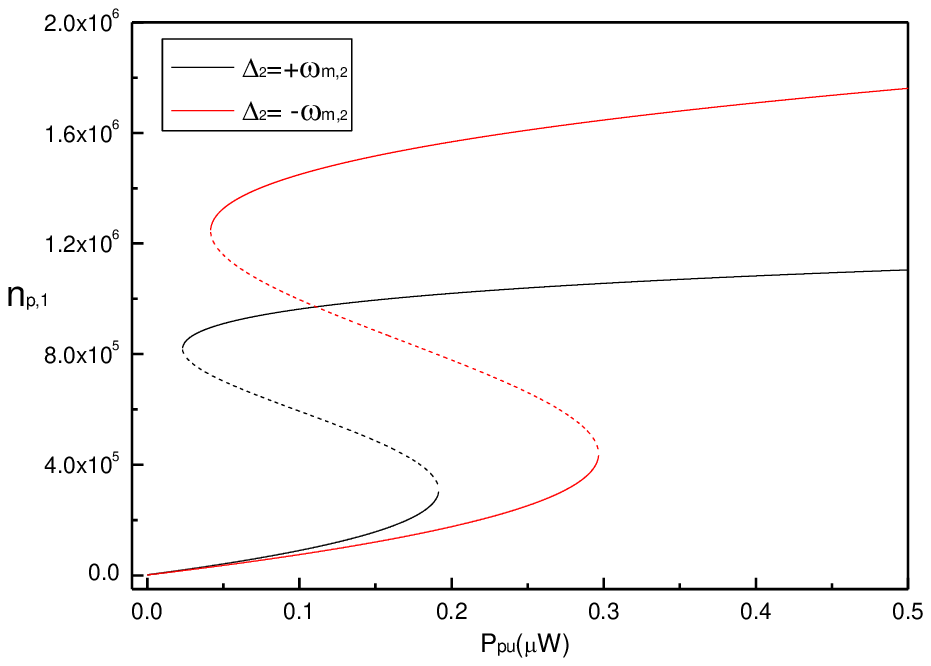}\caption{The mean
intracavity photon number in the left cavity $n_{p,1}$ as a function of
$P_{pu}$ with $P_{co}=0.010\mu$W, $J=$ $2\pi \times0.09$GHz, $g_{11}=2\pi
\times850$kHz, $g_{12}=2\pi \times860$kHz, $g_{21}=2\pi \times400$kHz,
$g_{22}=2\pi \times405$kHz, and $\Delta_{1}=2\pi \times2.0$GHz for different
values of $\Delta_{2}$: $\Delta_{2}=\omega_{m,2}$ and $\Delta_{2}%
=-\omega_{m,2}$, respectively.}%
\end{figure}

\begin{figure}[ptb]
\label{Figure8}
\centering \includegraphics[width=9cm]{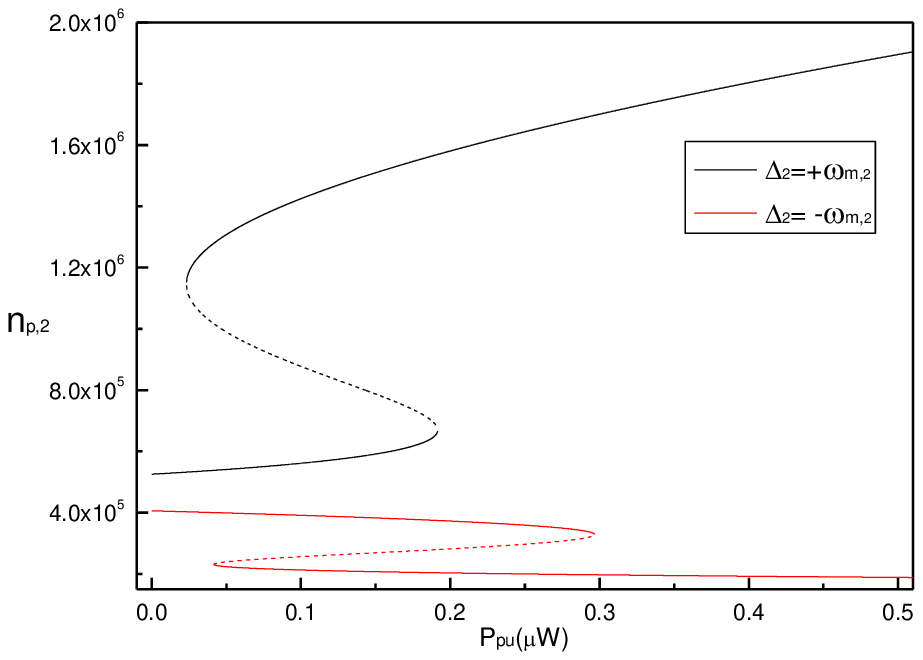}\caption{The mean
intracavity photon number in the right cavity $n_{p,2}$ as a function of
$P_{pu}$ with $P_{co}=0.010\mu$W, $J=$ $2\pi \times0.09$GHz, $g_{11}=2\pi
\times850$kHz, $g_{12}=2\pi \times860$kHz, $g_{21}=2\pi \times400$kHz,
$g_{22}=2\pi \times405$kHz, and $\Delta_{1}=2\pi \times2.0$GHz for different
values of $\Delta_{2}$: $\Delta_{2}=\omega_{m,2}$ and $\Delta_{2}%
=-\omega_{m,2}$, respectively.}%
\end{figure}

In this section, we will analyze the double bistability of mean intracavity
photon number in the right cavity under different coupling rate between the
left cavity and the first resonator. i.e., $g_{11}$. As shown in Fig.9, the
variation of coupling rate will leads to the modulation of threshold value in
the observation of bistability. Specifically, as the increase of coupling
rate, the first critical value of pump power for observing the first bistable
behavior is becoming lower; however, the second critical value of pump power
for the emergence of bistability is becoming larger with the strengthening of
coupling rate.

\begin{figure}[ptb]
\label{Figure9}
\centering \includegraphics[width=9cm]{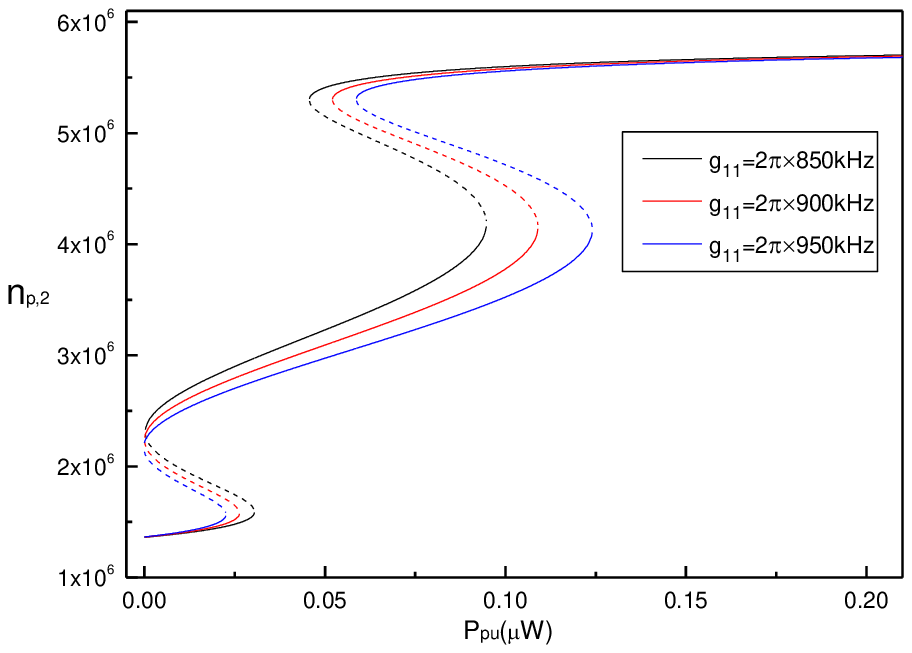}\caption{The mean
intracavity photon number in the right cavity $n_{p,2}$ as a function of
$P_{pu}$ with $P_{co}=0.010\mu$W, $J=$ $2\pi \times0.09$GHz, $g_{12}=2\pi
\times860$kHz, $g_{21}=2\pi \times400$kHz, $g_{22}=2\pi \times405$kHz, and
$\Delta_{1}=\Delta_{2}=2\pi \times2.0$GHz for different values of $g_{11}$:
$g_{11}=2\pi \times850$kHz, $2\pi \times900$kHz, and $2\pi \times950$kHz,
respectively.}%
\end{figure}

\subsection{Steady-state position of mechanical resonator}

It is shown in Eq.(\ref{6}) that the two steady-state positions of mechanical
resonator, i.e., $X_{1,s}$ and $X_{2,s}$, are directly related to mean
intracavity photon number $n_{p,1}$ and $n_{p,2}$. Thus it can be infer that
the steady-state positions should exhibit similar bistable behavior to the
mean intracavity photon number. According to the above analysis in the
bistable property of intracavity photon number, it is found that the large
photon number is necessary with the objective to realize the bistable
phenomenon; however, it can be seen from Fig.10 that the double bistability
can also be observed in relatively smaller photon number. For further
illustration, in Fig.10, the two steady-state positions of mechanical
resonator changing versus pump field driving strength $P_{pu}$ for different
control beam strength are presented, respectively. Initially, the steady-state
positions display bistability; however, when the control beam power is
increased to $P_{co}=0.02$ $\mu$W, then the double optical bistability in the
two steady-state positions of mechanical resonator can be observed. The
increasing control field power will result in stronger nonlineary effect on
photon number in both cavities, and then the radiation pressure on mechanical
resonators is risen, therefore the double optical bistability can be taken
place in the two steady-state positions of mechanical resonator.

\begin{figure}[ptb]
\label{Figure10}
\centering \includegraphics[width=9cm]{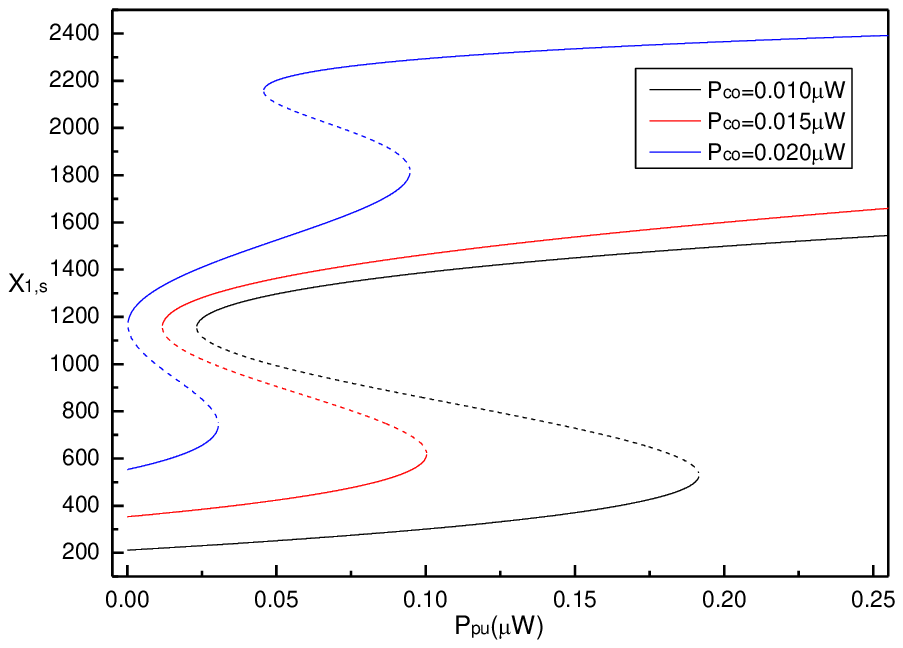}\caption{The mechanical
steady-state position $X_{1,s}$ as a function of $P_{pu}$ with $J=$
$2\pi \times0.09$GHz, $g_{11}=2\pi \times850$kHz, $g_{12}=2\pi \times860$kHz,
$g_{21}=2\pi \times400 $kHz, $g_{22}=2\pi \times405$kHz, and $\Delta_{1}%
=\Delta_{2}=2\pi \times2.0$GHz for different values of $P_{co}$: $P_{co}%
=0.010\mu$W, $0.015\mu$W, and $0.020\mu$W, respectively. }%
\end{figure}

Bistability in the mechanical resonator steady-state positions also
sensitively affected by the coupling coefficient $J$ between two cavities, as
depicted in Fig.11. Specifically, it can be straight forward to see that the
two critical points to observe the double optical bistability become smaller
as the strengthening of coupling coefficient between the two cavities.

\begin{figure}[ptb]
\label{Figure11}
\centering \includegraphics[width=9cm]{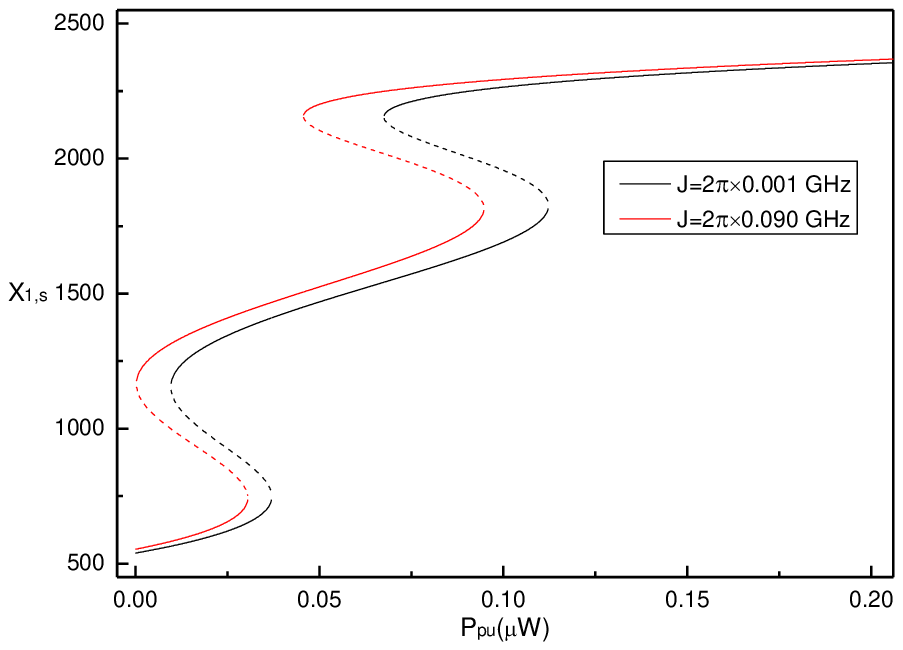}\caption{The first
mechanical steady-state position $X_{1,s}$ as a function of $P_{pu}$ with
$P_{co}=0.02$ $\mu$W, $g_{11}=2\pi \times850$kHz, $g_{12}=2\pi \times860$kHz,
$g_{21}=2\pi \times400$kHz, $g_{22}=2\pi \times405$kHz, $\Delta_{1}=\Delta
_{2}=2\pi \times2.0$GHz for different values of $J$: $J=2\pi \times0.001$GHz and
$2\pi \times0.090$GHz, respectively. }%
\end{figure}

As is shown in Fig.12(a), for a range of coupling rate $g_{11}$, the lower
driving strength is needed to observe the first bistable behavior in
steady-state position of the first mechanical resonator $X_{1,s}$ in the case
of larger coupling strength $g_{11}$, while the second threshold value to
observe the bistable phenomenon becomes larger in the case of larger coupling
strength $g_{11}$. Similar features have taken place in the steady-state
positions of the second mechanical resonator $X_{2,s}$, as shown in Fig.12(b),
the steady-state position of the second mechanical resonator $X_{2,s}$ versus
pump strength for several quantities in coupling strength $g_{11}$. Thus, the
changing of coupling rate between the first mechanical resonator and the left
cavity shall yields effective controllability over the behavior in the
steady-state position of the second mechanical resonator.

\begin{figure}[ptb]
\label{Figure12}
\centering \includegraphics[width=9cm]{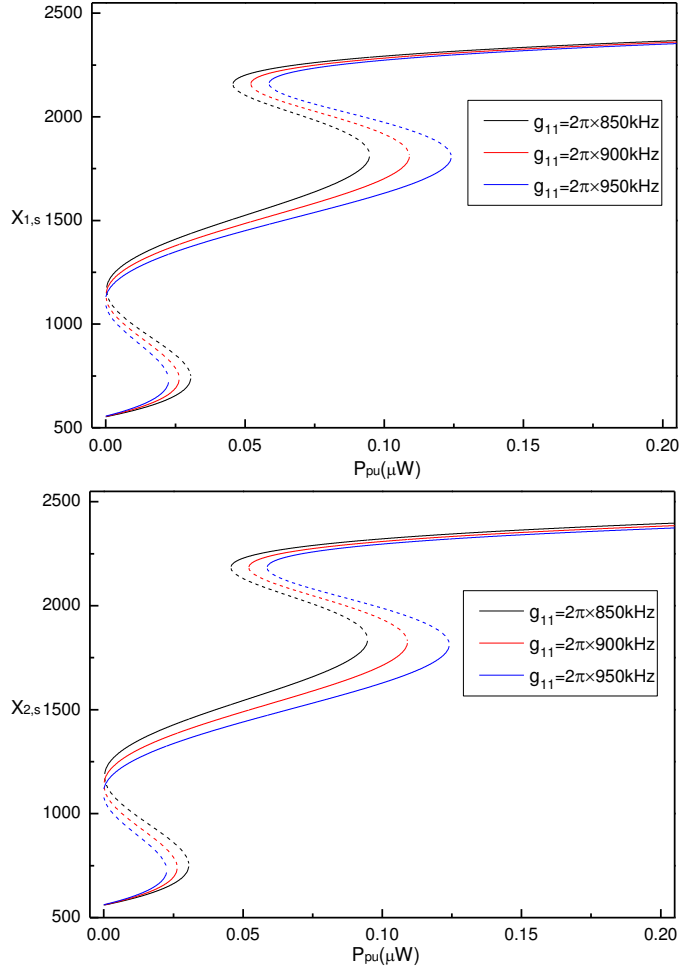}\caption{The first
mechanical steady-state position $X_{1,s}$ (a) and the second mechanical
steady-state positions $X_{2,s}$ (b) as a function of $P_{pu}$ with
$P_{co}=0.020\mu$W, $J=$ $2\pi \times0.09$GHz, $g_{12}=2\pi \times860$kHz,
$g_{21}=2\pi \times400$kHz, $g_{22}=2\pi \times405$kHz, and $\Delta_{1}%
=\Delta_{2}=2\pi \times2.0$GHz for different values of $g_{11}$: $g_{11}%
=2\pi \times850$kHz, $2\pi \times900$kHz, and $2\pi \times950$kHz, respectively.}%
\end{figure}

\section{Conclusions}

In summary, the controllability in optical bistability of intracavity photon
number in both cavities and the two steady-state positions of mechanical
resonators have been theoretically analyzed. It is shown that the
optomechanical system considered here enables robust controllability over the
bistable behavior of the intracavity photon number and steady-state positions
of mechanical resonators. Specifically, the bistable behavior of intracavity
photon number can be modulated by changing the control beam strength, coupling
coefficient between the two cavities, and coupling rates between cavities mode
and mechanical mode, respectively. It is proved that the bistable behavior of
the mean intracavity photon number in the right cavity can be tuned by
adjusting the strength of the pump laser beam driving the left cavity. And the
mean intracavity photon number is relatively larger in the red sideband regime
than that in the blue sideband regime. Moreover, we have shown the interesting
phenomenon that the double bistability of intracavity photon number in the
right cavity can be observed when the control field power increased to a
critical value. Besides, the critical values to observe bistability and double
bistability can be tuned by adjusting the coupling coefficient between two
cavities and the coupling rates between cavities mode and mechanical mode. The
larger coupling coefficient shall render smaller threshold value to observe
both bistability and double bistability. While in the coupling rates between
cavities mode and mechanical mode case, the larger coupling rates shall render
smaller threshold value to observe the first bistability and larger threshold
value to observe the second bistability. The controlling of optical
bistability shall have practical applications in building more efficient
all-optical switches and logic-gate devices for quantum computing and quantum
information processing.

\textbf{Acknowledgements}

The work was supported by the National Natural Science Foundation of China
(Grant Nos. 11874170, 11447202, 11447002, 11574295, and 11347026), Qing Lan
Project of Universities in Jiangsu Province; Postdoctoral Science Foundation
of China (2017M620593).

\end{document}